\documentclass[preprint]{aastex631}

\usepackage{hyperref}

\usepackage{amsmath,amsthm,latexsym}

\begin{document}

\title{Neutron star stability beyond the mass peak: assessing the role of out-of-equilibrium perturbations}

\author[0009-0002-7295-0470]{Martin O. Canullan-Pascual}
\affiliation{Grupo de Astrofísica de Remanentes Compactos\\ Facultad de Ciencias Astronómicas y Geofísicas, Universidad Nacional de La Plata\\ Paseo del Bosque S/N, La Plata (1900), Argentina}
\affiliation{CONICET, Godoy Cruz 2290, Buenos Aires (1425), Argentina}
\email{canullanmartin@fcaglp.unlp.edu.ar}

\author[0000-0002-2978-8079]{Germán Lugones}
\affiliation{Universidade Federal do ABC, Centro de Ciências Naturais e Humanas,\\ Avenida dos Estados 5001- Bangú, CEP 09210-580, Santo André, SP, Brazil.}
\email{german.lugones@ufabc.edu.br}

\author[0000-0001-8399-2567]{Milva G. Orsaria}
\affiliation{Grupo de Astrofísica de Remanentes Compactos\\ Facultad de Ciencias Astronómicas y Geofísicas, Universidad Nacional de La Plata\\ Paseo del Bosque S/N, La Plata (1900), Argentina}
\affiliation{CONICET, Godoy Cruz 2290, Buenos Aires (1425), Argentina}

\author[0000-0002-9071-4634]{Ignacio F. Ranea-Sandoval}
\affiliation{Grupo de Astrofísica de Remanentes Compactos\\ Facultad de Ciencias Astronómicas y Geofísicas, Universidad Nacional de La Plata\\ Paseo del Bosque S/N, La Plata (1900), Argentina}
\affiliation{CONICET, Godoy Cruz 2290, Buenos Aires (1425), Argentina}



\begin{abstract}
We investigate the radial stability of neutron stars under conditions where their composition may or may not remain in chemical equilibrium during oscillations. Using different equations of state that include nucleons, hyperons, and/or $\Delta$ resonances, we compute stellar configurations and examine their fundamental mode frequencies in two limiting scenarios. In one limit, nuclear reactions are fast enough to maintain chemical equilibrium throughout the pulsation, resulting in a lower effective adiabatic index, $\Gamma_{\mathrm{EQ}}$, and softer stellar responses. In the opposite limit, nuclear reactions are too slow to adjust particle abundances during oscillations, yielding a higher index, $\Gamma_{\mathrm{FR}}$, and stiffer stellar responses. We find that the equilibrium scenario triggers dynamic instability at the maximum mass configuration, whereas the frozen composition scenario allows stable solutions to persist beyond this mass, extending the stable branch. This effect is modest for simpler equations of state, but becomes increasingly pronounced for more complex compositions, where the emergence of new particle species at high densities leads to a significant disparity between $\Gamma_{\mathrm{EQ}}$ and $\Gamma_{\mathrm{FR}}$. Realistic conditions, in which different nuclear reactions have distinct timescales, will place the effective $\Gamma$ between these two extreme values. Short-timescale reactions push the star toward the equilibrium limit, potentially restricting the length of the stable branch. Conversely, slow reactions preserve a frozen composition, allowing the stable branch to grow. Thus, the actual extent of the stable configuration range depends critically on the interplay between nuclear-reaction timescales and the star’s fundamental oscillation period.
\end{abstract}



%\ifrs{HAY QUE UNIFICAR FORMATO DE REFERENCIAS}

%-------------------------------------------------------------
\section{Introduction} 
%-------------------------------------------------------------
\label{sec:intro}

Understanding the conditions under which neutron stars (NSs) remain dynamically stable is central to the study of compact objects.  Establishing stability criteria is crucial not only from a theoretical standpoint but also because modern observational techniques have provided stringent constraints on NS masses and radii. 
In particular, precise mass measurements of pulsars exceeding $2\,M_\odot$ \citep{Demorest:2010sdm, Antoniadis:2013amp, Arzoumanian:2018tny, Cromartie:2020rsd, Fonseca:2021rfa}, as well as the gravitational-wave event GW170817 and its electromagnetic counterpart \citep{Abbott:2017gwa, Abbott:2017oog, Abbott:2018exr, Abbott:2020goo}, have offered valuable insights into the equation of state (EoS) and the internal structure of NSs. Furthermore, observations from the NICER telescope \citep{Miller:2019pjm, Riley:2019anv, Miller:2021tro, Riley:2021anv} have refined radius measurements of isolated NSs, enabling direct comparisons between theoretical models and empirical data. As these increasingly precise measurements become available, it is now feasible to test stability predictions against astrophysical observations.
Determining the extent to which out-of-equilibrium processes can influence NS stability is therefore of vital importance.

The idea that radial pulsations might occur out of chemical equilibrium, i.e. that weak interactions responsible for adjusting the stellar composition to density changes may proceed more slowly than or on timescales comparable to the pulsation period, is not a new concept. Soon after the relativistic radial oscillation equations were formulated by \citet{Chandrasekhar:1964zz}, early works considered the possibility that the adiabatic index could differ significantly when matter is not in chemical equilibrium \citep{Meltzer1966, Chanmugam1977}. Although these pioneering studies identified considerable differences between equilibrium and non-equilibrium adiabatic indices, such differences did not produce any evident modification to the standard stability criterion. According to this criterion, stable configurations are confined between stationary points in the mass–central-density diagram, and the introduction of non-equilibrium effects did not seem to extend the stable region of configurations.

A more significant breakthrough emerged with the work of \citet{Gourgoulhon1995}, who investigated NS stability near the maximum mass limit under conditions of ``frozen'' composition due to slow weak interactions. Their study revealed that when matter cannot rapidly return to chemical equilibrium during oscillations, stable configurations can exist at central densities exceeding the nominal maximum mass density. In other words, slow reaction rates allow some NS configurations to remain stable even beyond what would be the traditional endpoint of stability if the matter were fully equilibrated. Although the effect was moderate (a few percent increase in density compared to equilibrium-based scenarios), it hinted that the standard maximum mass criterion might not fully capture the complexity of the stability landscape when compositional changes lag behind the pulsation cycle.

Subsequent investigations on protoneutron stars corroborated that the conventional maximum mass criterion still identifies the onset of relativistic instability, even in scenarios involving thermal effects, neutrino trapping, and frozen compositions \citep{Gondek:1997rpa}. However, these studies also showed that the so-called minimum mass criterion does not universally apply. While it works reliably for hot, isentropic, neutrino-opaque protoneutron stars, it fails for those with hot, isothermal interiors. These results highlight that assumptions about the internal thermal structure and reaction timescales can qualitatively alter the stability conditions, indicating that non-equilibrium processes need careful consideration.

More recently, out-of-equilibrium oscillations have been explored in depth within the context of hybrid stars featuring quark cores. \citet*{Pereira:2017rmp} demonstrated that if the phase transition between hadronic and quark matter is sharp and the conversion timescale between phases is slow relative to the oscillation period, a slow stable hybrid star (SSHS) branch emerges. In these configurations, stability can extend far into regions where the gravitational mass decreases with increasing central density, defying the standard mass-peak-based stability criterion. Subsequent work has confirmed the existence and robustness of SSHSs for various hybrid EoS \citep{mariani:2019mhs,malfatti:2020dba, tonetto:2020dgm, rodriguez:2021hsw, curin:2021hsw, Goncalves:2022ios, Mariani:2022omh, Ranea:2022bou, lugones:2023ama, Ranea:2023auq, Ranea:2023cmr, Rau:2023tfo, Rau:2023neo, Rather:2024roo, Jimenez:2024htq}, and related analyses have been extended to hybrid models with slow conversion reactions but without sharp phase discontinuities \citep{Gosh:2024ero}.

In light of this progression, it is clear that compositional complexity and its interplay with reaction timescales can substantially reshape the NS stability landscape. The objective of this work is to explore how chemically non-equilibrated radial pulsations might influence the stability of purely hadronic neutron stars beyond the mass peak, when the EoS includes multiple species, ranging from nucleons and hyperons to $\Delta$ resonances. By doing so, we aim to assess how non-equilibrium processes, potentially more pronounced in complex compositions, alter the conventional stability criteria and determine whether this complexity can give rise to long extended stable branches that transcend the standard mass-peak stability boundary.

The paper is organized as follows. In Section~\ref{sec:eos}, we describe the EoS considered and the parametrizations used to construct the NS models. In Section~\ref{sec:rad_osc}, we summarize the procedure for analyzing radial pulsations and assessing dynamical stability. In Section~\ref{sec:adiabatic}, we briefly discuss the adiabatic index and relaxation times, considering both the equilibrium and frozen composition limits, which provide lower and upper bounds for the star’s response to density changes.  Finally, Section~\ref{sec:results} presents our main results, including the existence of stable NSs beyond the maximum mass peak, while Section~\ref{sec:conclu} offers a summary and discussion of the key findings.

%-------------------------------------------------------------
\section{Equation of state} 
%-------------------------------------------------------------
\label{sec:eos}

To describe the matter inside NSs at different density regimes, we employ the BPS-BBP EoS for the outer and inner crust \citep{Baym:1971nsm, Baym:1971tgs}, and the SW4L parametrization for the cores of these objects \citep{Spinella:2019hns, malfatti:2020dba, Celi:2024doh}.

The Lagrangian density used to describe the baryon matter in the cores of NSs is given by
\begin{eqnarray}
  \mathcal{L_B} &=& \sum\limits_B \overline\psi_B\bigl[\gamma_{\mu}
    (i\partial^{\mu}-g_{\omega B}\omega^{\mu} - g_{\phi
      B}\phi^{\mu}       \nonumber\\ 
&&  -\tfrac{1}{2}g_{\rho B}(n_b)\boldsymbol{\tau}\cdot
    \boldsymbol{\rho}^{\mu})     -(m_B-g_{\sigma B} \sigma-g_{\sigma^* B}\sigma^*)\bigr]\psi_B   \nonumber\\
&&   +\tfrac{1}{2}\left(\partial_{\mu}\sigma\partial^{\mu}\sigma -m^2_{\sigma}\sigma^2\right)  + \tfrac{1}{2}\left(\partial_{\mu} \sigma^*\partial^{\mu}\sigma^* -m^2_{\sigma^*}\sigma^{*2}\right) \nonumber\\ 
&&  -\tfrac{1}{3}b_{\sigma}m_n\left(g_{\sigma N}\sigma\right)^3 -\tfrac{1}{4}c_{\sigma}\left(g_{\sigma
    N}\sigma\right)^4\nonumber\\ 
&&-\tfrac{1}{4}\omega_{\mu\nu}\omega^{\mu\nu} +\tfrac{1}{2}m^2_{\omega}\omega_{\mu}\omega^{\mu}   -\tfrac{1}{4}\phi^{\mu\nu}\phi_{\mu\nu}+\tfrac{1}{2}m^2_{\phi}\phi_{\mu} \phi^{\mu}       \nonumber\\ 
&&  -\tfrac{1}{4}\boldsymbol{\rho}_{\mu\nu}\cdot\boldsymbol{\rho}^{\mu\nu} +\tfrac{1}{2}m^2_{\rho}\boldsymbol{\rho}_{\mu} \cdot \boldsymbol{\rho}^{\mu}  ,
\label{Eq:laghad}
\end{eqnarray}
where the sum over $B$ includes the nucleons $N$, hyperons $H$ and the four states of the resonances $\Delta$, as indicated in Table~\ref{table:EoS_composition}.

%%%%%%%%%%%%%%%%%%%   TABLE 1
\begin{table}[!tb]
\centering
\begin{tabular}{c|cc|cc|cc}
\toprule   
Model     & $x_{\sigma \Delta}$ &  $x_{\omega \Delta}$ & $M_{\mathrm{max}}$ & $R_{\mathrm{max}}$ & $M_{\mathrm{T}}$ &   $R_{\mathrm{T}}$ \\ 
\toprule
 $N$                       & 0   &  0  & 2.33   &   11.4 & 2.33  &  11.4 \\    \hline
 $N\!\Delta1$       & 0.9  & 1.1  & 2.29   &   11.3 & 2.29  &  11.2 \\ 
 $N\!\Delta2$       & 1.25 & 1.1  & 2.20   &  10.3  & 2.21  &   10.3 \\ \hline
$N\!H\!\Delta1$   & 0.9 & 1.1 & 2.13   &   11.4  & 2.06  & 10.5 \\ 
$N\!H\!\Delta2$  & 1.25 & 1.1  & 2.14   &  10.5  & 2.08  &  9.90 \\ 
\toprule
\end{tabular}
\caption{Particle compositions of the EoS models used in this study ($N$ = nucleons, $H$ = hyperons, $\Delta$ = $\Delta$ resonances). All models also include electrons and muons. The table presents the coupling constant ratios $x_{\sigma \Delta}$ and $x_{\omega \Delta}$, which quantify the interaction strengths between $\Delta$ resonances and the $\sigma$ and $\omega$ meson fields, respectively. Additionally, it provides the maximum gravitational mass $M_{\mathrm{max}}$ with its corresponding radius $R_{\mathrm{max}}$, as well as the terminal mass $M_{\mathrm{T}}$ and its corresponding radius $R_{\mathrm{T}}$ for each model. The only difference between the models labeled 1 and 2 is the value of $x_{\sigma \Delta}$.}
\label{table:EoS_composition}
\end{table}

The Lagrangian density for the leptons is given by
\begin{eqnarray}
  \mathcal{L}_l =\bar{\psi}_l\left(i \gamma_\mu \partial^\mu-m_l\right) \psi_l , 
\end{eqnarray}
where $l= e^{-},\mu^{-}$.
The interactions between the baryons in Eq.~\eqref{Eq:laghad} are modeled in terms of scalar ($\sigma, ~\sigma^*$), vector ($\omega,~ \phi$), and isovector ($\rho$) meson fields. To determine the meson-nucleon and meson-hyperon coupling constants, $g_{iB}$, where $i=\sigma$, $\omega$, $\rho$, $\sigma^*$ and $\phi$ we use the Nijmegen extended soft-core (ESC08) model, based on a modified SU(3) symmetry \citep[details about the determination of the coupling constants for SW4L parametrization can be found in][]{malfatti:2020dba, Celi:2024doh}.
In particular, the quantities {$g_{\rho B}(n_b)$} in Eq.~\eqref{Eq:laghad}, denote density-dependent isovector meson--baryon coupling constants given by
\begin{equation}
g_{\rho B}(n_b) = g_{\rho B}(n_0)\,\mathrm{exp}\left[\,-a_{\rho}
  \left(\frac{n_b}{n_0} - 1\right)\,\right] \, ,
\end{equation}
where $n_b = \sum_B n_B$ is the total baryon number density.
The values of the constants involved are $a_{\rho}= 4.06 \times 10^{-3}$ and the nuclear saturation density, $n_0=0.15$ $\mathrm{fm}^{-3}$.

It is worth noting at this point that determining the interaction of mesons with $\Delta$-particles poses challenges due to limited experimental data. Consequently, since meson-$\Delta$ couplings are poorly constrained, specific coupling sets can be selected depending on the focus of the proposed study. For example, there are studies that vary the ratio $x_{i\Delta} \equiv g_{i\Delta}/g_{iN}$ within defined ranges (see, for example, \cite{Kalita:2024pti} and references therein) or propose ranges for these coupling ratios (see \cite{Sedrakian:2022dra} and references therein).

For the particle composition including $\Delta$-matter, we first explore the $(x_{\sigma \Delta}, x_{\omega \Delta})$ space over a broader range, considering the suggested ranges for the vector meson-$\Delta$ couplings from \cite{Sedrakian:2022dra}. Based on this exploration, we then select the combinations of $x_{\sigma \Delta}$ and $x_{\omega \Delta}$ as provided in Table~\ref{table:EoS_composition}.
It is important to note that in these cases, we used for the rest of the meson-$\Delta$ couplings the following values: $x_{\rho \Delta}=x_{\phi \Delta}=1.0$ and $x_{\sigma^* \Delta}=0.0$, where $x_{\sigma^*B} \equiv g_{\sigma^*B}/g_{\sigma^*\Lambda}$, and $g_{\sigma^* \Lambda} = 1.9242$.

To solve the equations of motion associated with Eq.~\eqref{Eq:laghad}, we use a density-dependent relativistic mean-field approximation. The corresponding coupled non-linear equations of motion are
\begin{eqnarray}
m_{\sigma}^2 \bar{\sigma} &=& \sum_{B} g_{\sigma B} n_B^s - \tilde{b}_{\sigma}  m_N   g_{\sigma N} (g_{\sigma N} \bar{\sigma})^2    \\
  &&  - \tilde{c}_{\sigma} g_{\sigma N}  (g_{\sigma N} \bar{\sigma})^3  \nonumber \\ 
m_{\sigma^*}^2 \bar{\sigma}^* &=& \sum_{B} g_{\sigma^* B} n_B^s\, , \\ 
m_{\omega}^2 \bar{\omega}     &=& \sum_{B} g_{\omega B} n_{B}\, , \\ 
m_{\rho}^2\bar{\rho}          &=& \sum_{B}g_{\rho B}(n_b)I_{3B} n_{B} \, , \\ 
m_{\phi}^2  \bar{\phi} &=& \sum_{B} g_{\phi B} n_{B}\, , 
\end{eqnarray}
where $\bar{\sigma}$, $\bar{\sigma}^*$, $\bar{\omega}$, $\bar{\rho}$, and $\bar{\phi}$ are the mean field values of the corresponding meson fields.  Here $I_{3B}$ is the 3-component of isospin, and $n_{B}^s$ and $n_{B}$ are the scalar and particle number densities for each baryon
$B$, which are given by
\begin{eqnarray}
n_{B}^s&=& \frac{1}{4\pi^2} \int^{p_{F_B}}_0 \frac{d^3p}{(2 \pi)^3}
\frac{m_B^*}{\sqrt{p^2+m_B^{*2}}}, \\ 
n_{B}&=& \frac{p_{F_B}^3}{ 3 \pi^2 },
\end{eqnarray}
where $m_B^*= m_B - g_{\sigma B}\bar{\sigma}-g_{\sigma^* B}\bar{\sigma}^*$ is the effective baryon mass and $p_{F_B}$ is the Fermi momentum.

The chemical potential of a baryon within the SW4L parametrization can be expressed as
\begin{equation}
\mu_B = g_{\omega B} \bar{\omega} + g_{\rho B} \bar{\rho} I_{3B}
+g_{\phi B} \bar{\phi} +\sqrt{p^2_{F_B}+m_B^{*2}}+ \widetilde{R} \, ,
\end{equation}
where the term $\widetilde{R} = \sum_B [\partial g_{\rho B}(n_b)/\partial n_b] I_{3B} n_B \bar{\rho}$,  is the rearrangement term necessary to guarantee thermodynamic consistency \citep{Hofmann:2001aot}.

The hadronic contribution to the pressure is given by
\begin{eqnarray}
P_h &=& \frac{1}{\pi^2}\sum_B \int^{p_{F_B}}_0 \! dp \,  \frac{p^4}{\sqrt{p^2+m_B^{*2}}}-\tfrac{1}{2} m_{\sigma}^2  \bar{\sigma}^2 \nonumber\\ 
& & - \tfrac{1}{2} m_{\sigma^*}^2 \bar{\sigma}^{* 2} + \tfrac{1}{2} m_{\omega}^2 \bar{\omega}^2 + \tfrac{1}{2} m_{\rho}^2 \bar{\rho}^2+ \tfrac{1}{2} m_{\phi}^2 \bar{\phi}^2\\ 
& & - \tfrac{1}{3} \tilde{b}_{\sigma} m_N (g_{\sigma N} \bar{\sigma})^3 - \tfrac{1}{4} \tilde{c}_{\sigma} (g_{\sigma N} \bar{\sigma})^4 + n_b \widetilde{R}.  \nonumber
\label{Eq:pressure}
\end{eqnarray}
To obtain the total pressure, $P$, we must include the contribution of the leptons $l$, which are modeled as a free degenerate Fermi gas of electrons and muons. Finally, the energy density of the system follows from
\begin{equation}
\epsilon = - P + \sum_{i=B,l} \mu_i \, n_i \, .
\label{eq:EoS}
\end{equation}

In this work, we focus on describing radial perturbations in spherically symmetric stars. We assume that the star being perturbed is initially in hydrostatic and thermodynamic equilibrium. The perturbations we consider cause slight deviations from this equilibrium configuration, temporarily shifting the matter into a new state, which may either remain in equilibrium or move out of equilibrium, depending on the rate of the chemical reactions involved. However, it is important to emphasize that the initial configuration is indeed in thermodynamic equilibrium. Therefore, when calculating the EoS of unperturbed matter, we adopt the standard conditions of chemical equilibrium under weak interactions. Since we are analyzing cold NSs, neutrinos can escape the system because their mean free path is much larger than the star's radius, implying that the neutrino chemical potential is zero ($\mu_{\nu_e} = 0$). Finally, we assume local electric charge neutrality and baryon-number conservation. In this context, the chemical potentials of unperturbed matter are related by the following set of equations
\begin{eqnarray}
\mu_B &=& \mu_n + q_B\,\mu_e, \label{chempot} \\
\sum_{i=B,l} q_i n_i &=& 0, \\
n_b - \sum_B n_B &=& 0,
\end{eqnarray}
where $q_B$ ($q_l$) is the baryon (lepton) electric charge, and $\mu_n$ and $\mu_e$ are the chemical potentials of neutrons and electrons, respectively.

%-------------------------------------------------------------
\section{Radial Oscillations and dynamical stability} 
%-------------------------------------------------------------
\label{sec:rad_osc}

In the following, we outline the procedure for analyzing the dynamical stability of non-rotating, spherically symmetric stellar models. The space-time metric for such configurations is expressed as
\begin{equation}
ds^2 = e^\nu dt^2 - e^\lambda dr^2 - r^2 (d\theta^2 + \sin^2\theta \, d\phi^2),
\end{equation}
where $\nu$ and $\lambda$ are functions of the radial coordinate $r$. The hydrostatic equilibrium of these configurations is described by the Tolman-Oppenheimer-Volkoff (TOV) equations, which can be solved once an EoS of the form $P = P(\epsilon)$ is specified.

To analyze the response of the star to small disturbances, we introduce into Einstein's equations a field of small radial Lagrangian displacements of fluid elements, which induces small changes in the metric and thermodynamic quantities. Working within the linear regime and assuming a harmonic time dependence for perturbations of the form $\exp(i \tilde{\omega} t)$, the pulsation equations are reduced to a Sturm-Liouville problem for the relative radial displacement $\xi \equiv \Delta r / r$, the Lagrangian perturbation in pressure $\Delta P$, and the associated eigenfrequencies $\omega = \tilde{\omega} / (2\pi)$. 
In this study, we employ the equations detailed in Appendix~\ref{sec:appendix}:
\begin{eqnarray}
\frac{d\xi}{dr}&=&-\frac{1}{r}\bigg(3\xi+\frac{\Delta P}{\Gamma P}\bigg)-\frac{dP}{dr}\frac{\xi}{(P+\epsilon)},
\label{Eq:ecuacionparaXI}
\end{eqnarray}
\begin{eqnarray}
\label{Eq:ecuacionparaP}
\frac{d\Delta P}{dr}&=&\xi \bigg\{{\tilde{\omega}}^{2}e^{\lambda-\nu}(P+\epsilon)r-4\frac{dP}{dr} \bigg \} \nonumber  \\
&+&\xi\bigg \{\bigg(\frac{dP}{dr}\bigg)^{2}\frac{r}{(P+\epsilon)}-8\pi e^{\lambda}(P+\epsilon)Pr \bigg \} \\
&+&\Delta P\bigg \{\frac{dP}{dr}\frac{1}{(P+\epsilon)}-
4\pi(P+\epsilon)r e^{\lambda}\bigg\}\nonumber ,
\end{eqnarray}
where $\Gamma$ is the adiabatic index, which will be examined in detail in the next section.

To solve Eqs. (\ref{Eq:ecuacionparaXI}) and (\ref{Eq:ecuacionparaP}), two boundary conditions are required. The condition of regularity at $r=0$ requires that
\begin{equation}\label{Eq:DeltaP}
(\Delta P)_{r=0}=-3(\xi \Gamma P)_{r=0}.
\end{equation}
Notice that, as we are working with linearized equations, the amplitude of the perturbation is irrelevant. For this reason, the eigenfunctions can be normalized so that $\xi(0)=1$.  The radius of the star, $R$, is determined by imposing the condition of vanishing pressure at the surface, $P(R)=0$, when solving the TOV equations. For the pressure perturbation, the boundary condition required at the surface is that $\Delta P$ must vanish:
\begin{equation}\label{PenSuperficie}
(\Delta P)_{r=R}=0.
\end{equation}

Since this system of differential equations and boundary conditions constitutes a Sturm-Liouville problem, the real eigenvalues $\omega_n^2$ are ordered and the fundamental mode, characterized by $\omega_0^2$, determines the stability of the star. A positive $\omega_0^2$ indicates stable oscillations, whereas a negative $\omega_0^2$ signals exponential growth of the perturbation and stellar instability.

Calculations using a wide variety of EoS agree that NSs with small central baryon number density $n_{b_c}$ are dynamically stable. As we move along the curve $M$ versus $R$ in the direction of increasing $n_{b_c}$, the values of $\omega_0^2$ initially increase. However, beyond a certain point, they begin to decrease until they finally become zero at the last stable object in that sequence. Under the condition that perturbations occur in chemical equilibrium (cold-catalyzed matter), the last dynamically stable object in the sequence —where $\omega_0^2 = 0$— is the one with the maximum mass. Beyond this point, any additional mass leads to instability, causing the star to disrupt or collapse into a black hole.

However, this is not the case in more general situations. When perturbations are considered that do not maintain chemical equilibrium -such as when the reactions restoring equilibrium are slow compared to the oscillation timescale - the static stability criterion changes. In these scenarios, the point at which $\omega_0^2 = 0$ does not necessarily occur at the maximum mass configuration. In analogy to the definition introduced in \citet*{Pereira:2017rmp}, we will refer to the star with $\omega_0^2 = 0$ as the \textit{terminal} configuration.

%-------------------------------------------------------------
\section{Adiabatic index and relaxation times} 
%-------------------------------------------------------------
\label{sec:adiabatic}

The adiabatic index introduced in the previous section is defined as
\begin{equation}
\Gamma  = \frac{n_b}{P} \frac{\partial P}{\partial n_b}  \bigg|_{ \{ \mathcal{C}\} },
\label{Eq:Gamma_def}
\end{equation}
and measures how pressure changes with local baryon number density under the set of physical conditions $\{ \mathcal{C} \}$ that characterize the perturbation. 
To understand which conditions $\{ \mathcal{C} \}$ should be applied, it is necessary to consider the typical timescales $\tau_\mathrm{nuc}$ of the various nuclear reactions that occur within the NS and compare them with the pulsation period $\mathcal{T}$ \citep{Meltzer1966}, which is on the order of 0.1 to 1 millisecond. As emphasized by \cite{Haensel:2002aio}, equilibration processes in NSs occur on vastly different timescales relative to $\mathcal{T}$   \citep[also see the recent work of][where this is extended to hybrid stars]{Rau:2023neo}. Strong and electromagnetic elastic collisions, as well as strong interaction processes that conserve strangeness, have extremely short relaxation times (on the order of $10^{-16}$ to $10^{-19}$ seconds), ensuring immediate equilibration without changing particle fractions. In contrast, weak interaction processes operate much more slowly. For a typical temperature of $10^9\,\text{K}$ (compatible with the zero-temperature approach of our study), modified Urca processes are the slowest, with relaxation times of several days, while direct Urca processes are faster, but still take several seconds. Nonleptonic processes that change strangeness are the fastest among the weak interactions, with relaxation times of about 1 millisecond, playing a crucial role in equilibrating hyperonic matter. It is important to note that these rough estimates are valid for $\delta \mu \ll T$, where $\delta \mu$ is a typical pulsation amplitude of chemical potentials. In the regime where $\delta \mu \gtrsim T$, these estimates may change significantly.

We can bypass the complexities of calculating the reaction rates discussed earlier by focusing on two limiting cases: oscillations in chemical equilibrium ($\tau_\mathrm{nuc} \ll \mathcal{T}$) and oscillations with frozen composition ($\tau_\mathrm{nuc} \gg \mathcal{T}$). As noted in the previous discussion, the actual scenario does not align fully with either of these extremes, as some reactions occur faster than $\mathcal{T}$, while others are slower. However, analyzing these limiting cases will allow us to place constraints on the dynamical stability of stellar models in the mass-radius diagram. 

If $\tau_\mathrm{nuc} \ll \mathcal{T}$, the perturbed matter has enough time to adjust its composition through nuclear reactions, which means that it will oscillate in a state of permanent chemical equilibrium. Therefore, the adiabatic index for this scenario can be directly determined from the EoS used to calculate the TOV equations, 
\begin{equation}
\Gamma_{\mathrm{EQ}} = \frac{n_b}{P} \frac{\partial P}{\partial n_b} \bigg|_\mathrm{ \{EoS\} } . 
\label{Eq:Gammaeq}
\end{equation}
In contrast, if $\tau_\mathrm{nuc} \gg \mathcal{T}$, nuclear reactions cannot adjust the abundances of particles during an oscillation period, and the relative fractions of different species of particles remain fixed as the density changes. This situation is referred to as a perturbation occurring with ``frozen'' chemical composition. Under these conditions, the adiabatic index $\Gamma_\mathrm{FR}$ must be determined by taking the derivative while keeping the abundances $Y_i=\frac{n_i}{n_b}$ of all particle species constant:
\begin{equation}
\Gamma_\mathrm{FR} = \frac{n_{b}}{P}\left.\frac{\partial P}{\partial n_b}\right|_{\{Y_i\}}  .
\label{Eq:Gammaf}
\end{equation}
Since the unperturbed model is locally electrically neutral, the condition of frozen abundances automatically ensures baryon number conservation and electrical charge neutrality of the perturbed matter as well (see \citet{lugones:1998eot, Benvenuto:1999tpt}, though in a slightly different context).

The adiabatic indices defined earlier represent the lower and upper bounds of $\Gamma$. In equilibrium, the system can fully adjust its composition through nuclear reactions, reducing the pressure response to a change in density by redistributing particles among different species. In contrast, when the composition is frozen, the particles cannot be redistributed at all, leading to a higher pressure than in equilibrium for the same density change. As a result, $\Gamma_\mathrm{FR}$ is always greater than $\Gamma_\mathrm{EQ}$. In a more realistic scenario, where some redistribution of particles is possible but not complete, the adiabatic index will fall between $\Gamma_\mathrm{EQ}$ and $\Gamma_\mathrm{FR}$.

%%%%%%%%    FIGURE 1
\begin{figure*}
\includegraphics[scale=0.35]{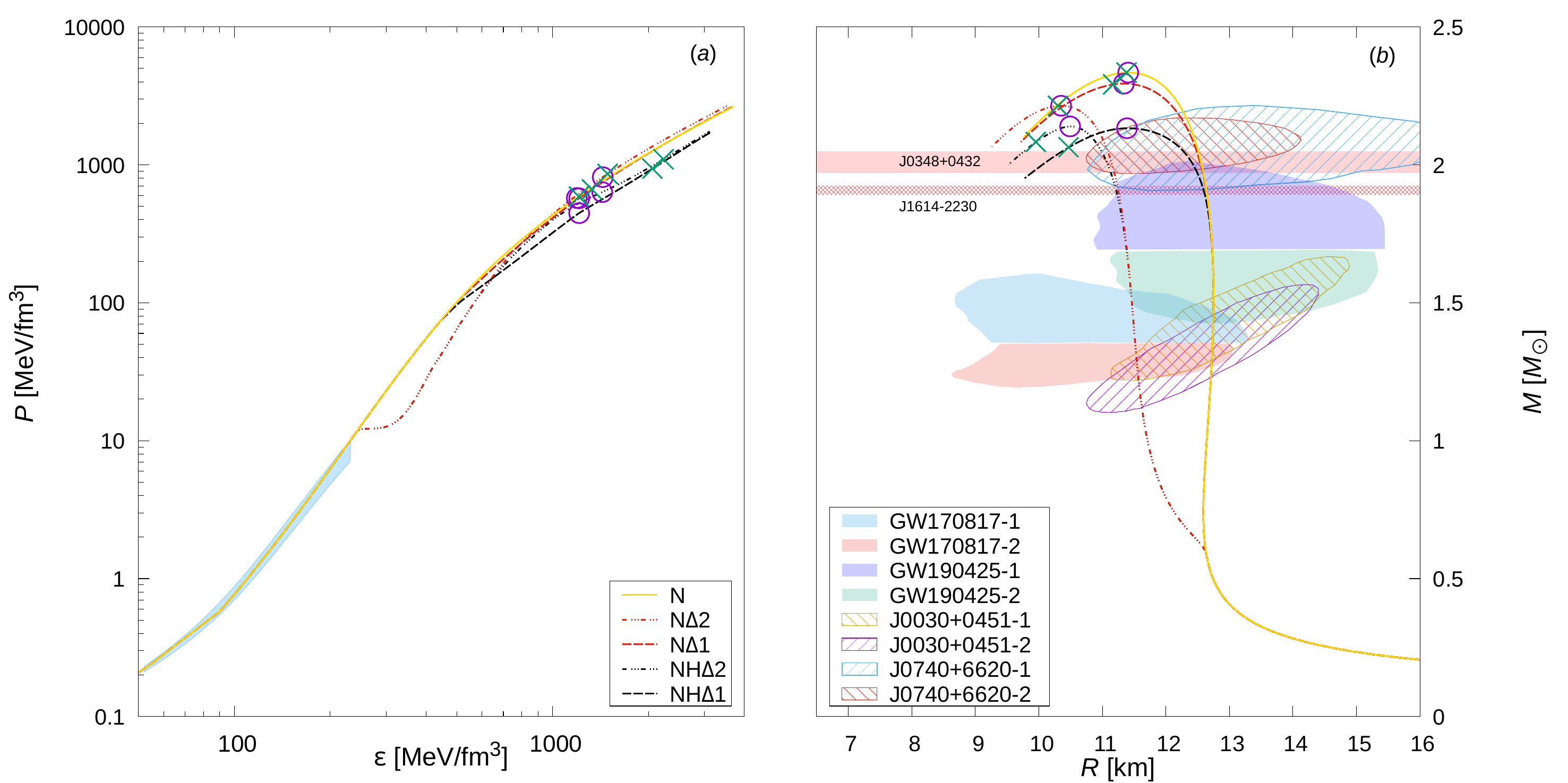}
\caption{(a) Pressure $P$ against the energy density $\epsilon$ for the baryonic compositions described in the text: $N$, $N\!\Delta$, and $N\!H\!\Delta$ for two parameterizations of $x_{\sigma\Delta}$. Our models satisfy the chiral EFT constraint (light blue area) from \citet{Drischler:2021lma}. (b) Mass-radius relationships for NSs based on the EoS shown in (a), compared with observational constraints from pulsars and gravitational wave events. The purple circles represent the maximum mass configurations, while the green crosses indicate the terminal mass configurations for the frozen composition case. }
\label{Fig:EoS_MR}
\end{figure*}

%%%%%%%%    FIGURE 2
\begin{figure}
\begin{center}
\includegraphics[width=0.6\linewidth]{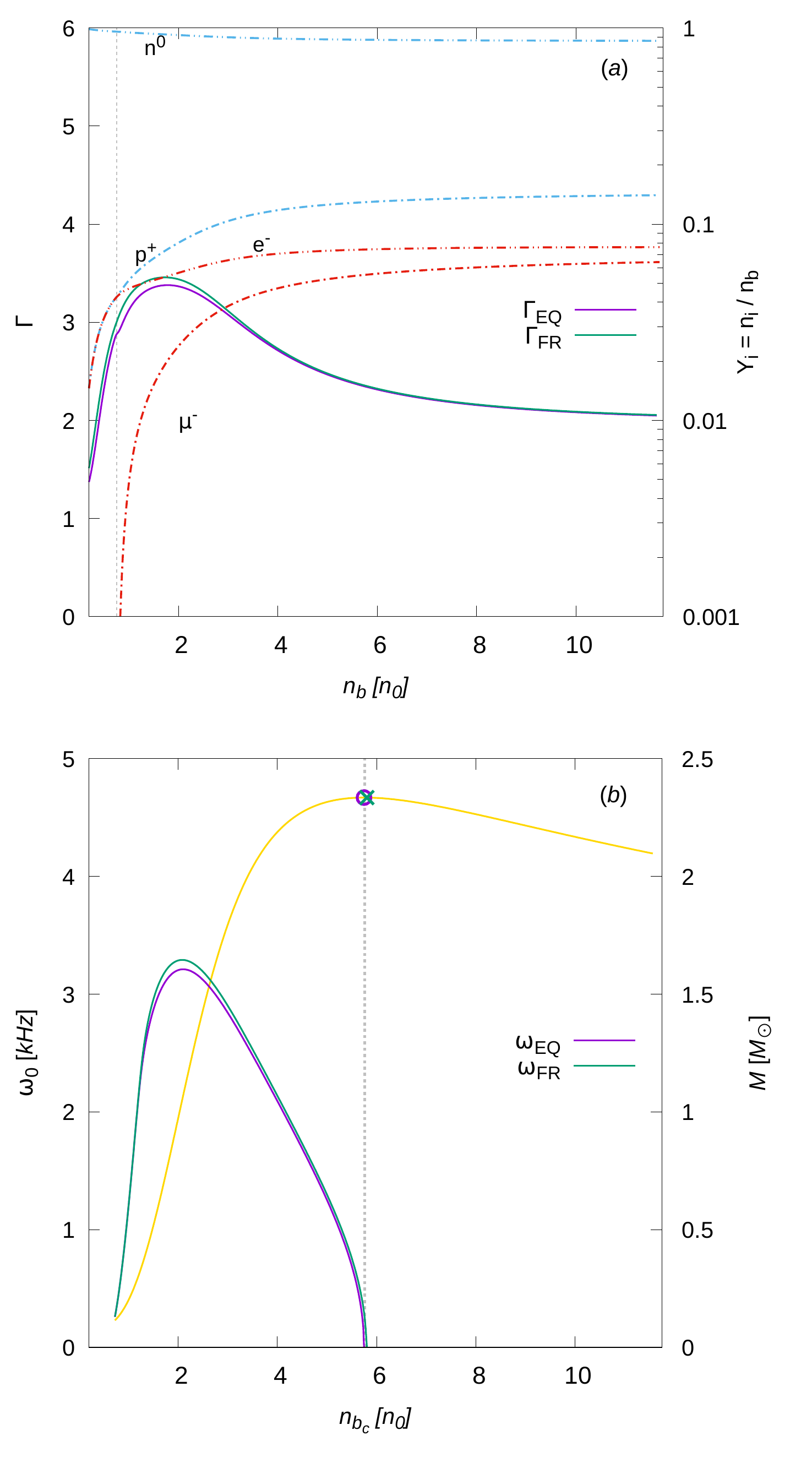}
\end{center}
\caption{(a) The adiabatic indices $\Gamma_{\mathrm{EQ}}$ and $\Gamma_{\mathrm{FR}}$ plotted against the baryon number density $n_b$ (in units of the nuclear saturation density $n_0$) for a NS composition consisting of protons, neutrons, electrons, and muons. The different particle abundances $Y_i = n_i / n_b$ are shown as a function of $n_b$. (b) The fundamental radial oscillation frequency $\omega_0$ (in kHz) as a function of the central baryon number density $n_{b_c}$ (in units of $n_0$) and the corresponding NS mass $M$. The results for $\omega_0$ are shown for oscillations in chemical equilibrium ($\omega_{\mathrm{EQ}}$) and with a frozen composition ($\omega_{\mathrm{FR}}$). }  
\label{Fig:gamma-pop-M-nu}  
\end{figure}

%%%%%%%%    FIGURE 3
\begin{figure*}[tbh]
\centering
\includegraphics[width=\textwidth]{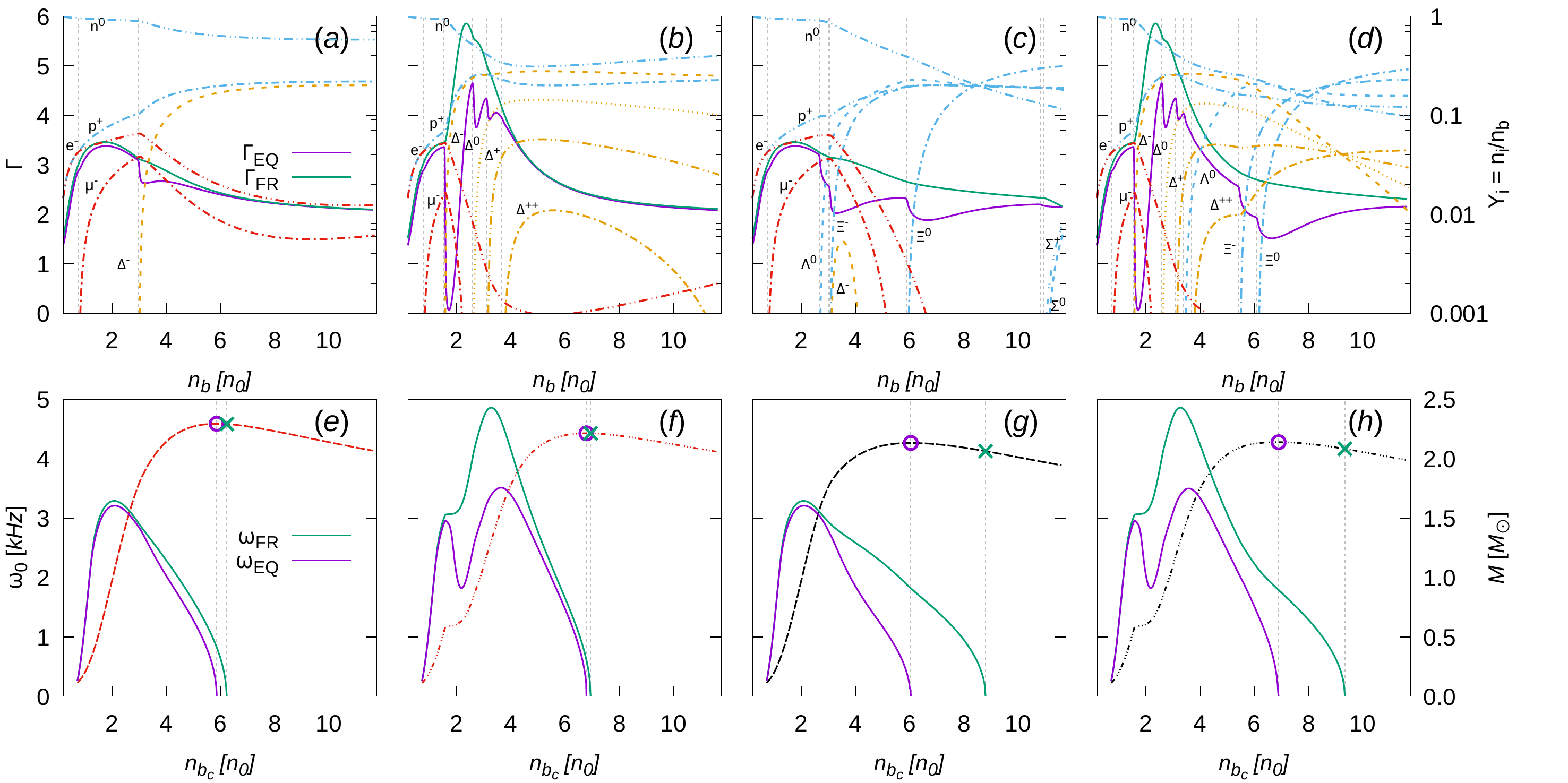}
\caption{Panels (a) and (b) display the particle populations and the adiabatic indices $\Gamma_{\mathrm{EQ}}$ and $\Gamma_{\mathrm{FR}}$ as functions of  $n_b$ for the $N\!\Delta$ case with $x_{\sigma \Delta}=0.9$ and $x_{\sigma \Delta}=1.25$, respectively. Panels (b) and (d) present similar analyses for the $N\!H\!\Delta$ case. The corresponding fundamental oscillation frequencies $\omega_{\mathrm{EQ}}$ and $\omega_{\mathrm{FR}}$ are plotted against the central baryon number density $n_{b_c}$ in panels (e), (f), (g), and (h), along with the gravitational mass curves as functions of $n_{b_c}$.    }   
\label{Fig:multiple-gamma-frec}
\end{figure*}

%-------------------------------------------------------------        
\section{Results} 
%-------------------------------------------------------------
\label{sec:results}

In this work, we consider three types of composition for the baryonic sector of the EoS, with the leptonic part always consisting of electrons and muons. The three cases, presented in Table~\ref{table:EoS_composition} and depicted in Figure~\ref{Fig:EoS_MR}(a), are as follows:
\begin{itemize}
    \item only protons and neutrons ($N$), represented by the solid yellow line;
    \item protons, neutrons, and $\Delta$ resonances ($N\!\Delta$), represented by red lines;
    \item the complete baryon octet plus $\Delta$ resonances ($N\!H\!\Delta$), shown by black lines.
\end{itemize}
For the cases involving $\Delta$ resonances, we adopt two different parameterizations of $x_{\sigma\Delta}$: specifically, $x_{\sigma\Delta}=0.9$ (dashed lines) and $x_{\sigma\Delta}=1.25$ (dashed-dotted lines). The resulting EoS are presented alongside the light blue area, which corresponds to the constraints from chiral EFT calculations performed by \citet{Drischler:2021lma}.

Using the aforementioned compositions, we solved the TOV equations to obtain stellar configurations in hydrostatic equilibrium. Figure~\ref{Fig:EoS_MR}(b) presents the mass-radius relationships for each EoS, using the same colors and line styles as in Figure~\ref{Fig:EoS_MR}(a). The resulting curves are consistent with the modern astronomical constraints discussed in Section~\ref{sec:intro}.
In every curve in Figure~\ref{Fig:EoS_MR}(b), two special configurations are highlighted: the maximum mass configuration (indicated by a purple circle) and the terminal configuration (defined in Section \ref{sec:rad_osc}) for the frozen composition case (marked by a green cross).
As previously emphasized, the maximum mass object represents the first unstable configuration in the sequence of stars when perturbations occur under permanent chemical equilibrium (i.e., when using $\Gamma_\mathrm{EQ}$ in the pulsation equations). In contrast, if the composition remains ``frozen" during pulsations (i.e., when using $\Gamma_\mathrm{FR}$), the first unstable configuration is identified by locating the point where $\omega_0^2 = 0$.

The results in Figure~\ref{Fig:EoS_MR}(b) lead to an intriguing conclusion: in the frozen composition scenario, the terminal configuration is located beyond the maximum mass configuration, resulting in an extended branch of stable configurations where $\partial M / \partial n_{b_c} < 0$. The length of this extended branch is relatively short for cases with less diversified compositions, such as those with only nucleons or with nucleons and $\Delta$ resonances, but is considerably longer for the $N\!H\!\Delta$ composition. In the following subsections, we analyze each of the models in detail.

%--------------------------------
\subsection{Model $N$}
%--------------------------------

In Figure~\ref{Fig:gamma-pop-M-nu}(a), we present the particle abundances along with the two limiting adiabatic indices, $\Gamma_{\mathrm{EQ}}$ and $\Gamma_{\mathrm{FR}}$, as functions of the baryon number density for the case $N$ in Table~\ref{table:EoS_composition}.

To draw physical conclusions about the behavior of $\Gamma_{\mathrm{EQ}}$ and $\Gamma_{\mathrm{FR}}$, it is instructive to analyze how the abundances of particles redistribute in response to variations in density, as shown by the abundances of particles $Y_i$ in panel (a) of the figure. In the region where $n_b \lesssim 2\,n_0$, the abundances of electrons and protons exhibit significant variations as the density changes. As the density increases to $\,n_0 \lesssim n_b \lesssim 3\,n_0$, even small changes in $n_b$ lead to substantial alterations in the abundance of muons. Consequently, in the case of perturbations occurring in chemical equilibrium, there will be considerable redistribution of particles, which mitigates the increase in pressure in response to changes in density. In this regime, $\Gamma_{\mathrm{EQ}}$ is markedly lower than $\Gamma_{\mathrm{FR}}$ due to this redistribution. At higher densities ($n_b \gtrsim 3\,n_0$), the differences between $\Gamma_{\mathrm{EQ}}$ and $\Gamma_{\mathrm{FR}}$ decrease substantially. This is because the curves representing the abundances of particles become more horizontal, indicating that small changes in $n_b$ do not result in significant particle redistribution. Thus, a perturbation in chemical equilibrium behaves similarly to one with a frozen composition since the particle abundances do not adjust substantially. Therefore, at these higher densities, both adiabatic indices tend to converge in their behavior.

In Figure~\ref{Fig:gamma-pop-M-nu}(b), the frequency of the fundamental mode is shown as a function of the central baryon number density, $n_{b_c}$, for both equilibrium ($\omega_{\mathrm{EQ}}$) and frozen composition ($\omega_{\mathrm{FR}}$) cases. These results were obtained by solving the radial oscillation equations, Eqs. \eqref{Eq:ecuacionparaXI} and \eqref{Eq:ecuacionparaP}. For comparison purposes, the NS mass is also displayed as a function of $n_{b_c}$. 

The fundamental oscillation frequency for both cases begins near zero at low $n_{b_c}$, increases to a maximum, and then gradually decreases as $n_{b_c}$ continues to rise, eventually reaching zero at the last stable configuration. This behavior is characteristic of NS models, which typically exhibit both a minimum mass and a maximum (or terminal) mass beyond which the star becomes dynamically unstable. In particular, $\omega_{\mathrm{FR}}$ is generally higher than $\omega_{\mathrm{EQ}}$, particularly at lower densities.  This difference arises because, with $\Gamma_{\mathrm{EQ}} < \Gamma_{\mathrm{FR}}$, the star exhibits a stiffer response to perturbations in the frozen composition case, leading to stronger restoring forces and thus higher oscillation frequencies.  At higher central densities ($n_{b_c} \gtrsim 3\,n_0$), the regime in which both adiabatic indices are nearly identical becomes dominant in the interior of the NS, leading to similar dynamical behavior in both the equilibrium and frozen composition scenarios. Consequently, $\omega_{\mathrm{EQ}}$ and $\omega_{\mathrm{FR}}$ tend to converge at these higher densities.

The yellow line in Figure~\ref{Fig:gamma-pop-M-nu}(b) shows the NS mass as a function of $n_{b_c}$. As in Figure~\ref{Fig:EoS_MR}, the maximum mass configuration is represented by a violet circle, while the terminal mass configuration is marked by a green cross. The point where $\omega_{\mathrm{EQ}} = 0$ corresponds to the maximum mass configuration. However, in the frozen composition scenario, instability occurs at a higher central density, where $\omega_{\mathrm{FR}} = 0$, indicating that stars can remain stable beyond the maximum mass limit. Although the difference between the maximum mass and the terminal mass is small in this case, we will show below that it can become significant depending on the composition of the matter.

%--------------------------------
\subsection{Model $N\!\Delta1$}
%--------------------------------

In Figure~\ref{Fig:multiple-gamma-frec}(a), we present the same analysis as in Figure~\ref{Fig:gamma-pop-M-nu}(a), but for the case $N\!\Delta$ with $x_{\sigma \Delta}=0.9$. For densities $n_b \lesssim 3\,n_0$, the abundances of $n$, $p^+$, $e^-$, and $\mu^-$, as well as the indices $\Gamma_{\mathrm{EQ}}$ and $\Gamma_{\mathrm{FR}}$, closely resemble those shown in Figure~\ref{Fig:gamma-pop-M-nu}(a). However, the appearance of $\Delta^-$ resonances at $n_b \sim 3\,n_0$ introduces an additional channel that enables the redistribution of other particles, as clearly seen in the kinks in the abundances around $n_b \sim 3\,n_0$. Consequently, a density perturbation that occurs in chemical equilibrium around $n_b \sim 3\,n_0$ leads to significant changes in the partial pressures of all particle species, resulting in a pronounced drop in $\Gamma_{\mathrm{EQ}}$. In contrast, $\Gamma_{\mathrm{FR}}$, as usual, exhibits a more gradual decline as density increases. At densities $n_b \gtrsim 5-6\,n_0$, both adiabatic indices tend to converge as the particle abundances stabilize.

The fundamental frequencies corresponding to the model in Figure~\ref{Fig:multiple-gamma-frec}(a) are shown in Figure~\ref{Fig:multiple-gamma-frec}(e).  Both $\omega_{\mathrm{FR}}$ and $\omega_{\mathrm{EQ}}$ exhibit qualitatively similar behavior to that seen in the simpler model of Figure~\ref{Fig:gamma-pop-M-nu}(b).  However, the main difference lies in the fact that, for central densities exceeding $3\,n_0$, $\omega_{\mathrm{FR}}$ is significantly higher than $\omega_{\mathrm{EQ}}$. This occurs because, for $n_{b_c} \gtrsim 3\,n_0$, the region where $\Gamma_{\mathrm{FR}}$ is much larger than $\Gamma_{\mathrm{EQ}}$ covers a substantial portion of the star. As a result, the overall restoring force is stronger in the frozen composition scenario, leading to a higher oscillation frequency.
Additionally, the extended branch where $\omega_{\mathrm{FR}} > 0$ stretches beyond the point where $\omega_{\mathrm{EQ}} = 0$, indicating that the star can remain dynamically stable at higher central densities compared to the equilibrium scenario.

%--------------------------------
\subsection{Model $N\!\Delta2$}
%--------------------------------

In Figure~\ref{Fig:multiple-gamma-frec}(b) we show the same as in Figure~\ref{Fig:multiple-gamma-frec}(a) but with the parameterization $x_{\sigma \Delta}=1.25$, which allows the appearance of all four types of resonances $\Delta$. The resonances $\Delta^-$, $\Delta^0$, $\Delta^+$, and $\Delta^{++}$ emerge successively at densities that are approximately $1.5\,n_0$, $2.5\,n_0$, $3\,n_0$, and $3.5\,n_0$, respectively. In each case, the corresponding abundance increases sharply, opening a new channel for the redistribution of the other particles, and a significant drop in $\Gamma_{\mathrm{EQ}}$ is observed due to substantial softening of the EoS.

The most dramatic effect is associated with the appearance of $\Delta^-$, which drives $\Gamma_{\mathrm{EQ}}$ to an almost vanishing value. This is related to the quasi-plateau observed in the EoS in Figure~\ref{Fig:EoS_MR}(a). The drops related to the appearance of $\Delta^0$, $\Delta^+$, and $\Delta^{++}$, while still significant, are comparatively less pronounced. As the density increases beyond $n_b \sim 4\,n_0$, where the abundances of all resonances $\Delta$ stabilize and begin to decrease, the $\Gamma_{\mathrm{EQ}}$ curve becomes nearly indistinguishable from the $\Gamma_{\mathrm{FR}}$ curve.

The behavior of $\omega_{\mathrm{EQ}}$ and $\omega_{\mathrm{FR}}$ shown {in Figure~\ref{Fig:multiple-gamma-frec}(f)} is similar to that previously described, except for the appearance of a dip in $\omega_{\mathrm{EQ}}$, which is associated with the softening introduced by the sequential appearance of the resonances $\Delta$ over a narrow range of densities. Despite these differences, the length of the extended branch remains small, as in the previous cases. This is because the main difference between $\Gamma_{\mathrm{FR}}$ and $\Gamma_{\mathrm{EQ}}$ is confined to a density range between $2\,n_0$ and $4\,n_0$, which affects the oscillation frequency of intermediate-mass stars, but not those with masses close to the maximum mass.

%--------------------------------
\subsection{Model $N\!H\!\Delta1$}
%--------------------------------

In Figure~\ref{Fig:multiple-gamma-frec}(c), we present the results for the $N\!H\!\Delta$ case with $x_{\sigma \Delta}=0.9$. The $\Lambda^0$ hyperon appears around $2.5\,n_0$, followed by the $\Xi^-$ and $\Delta^-$ resonances at approximately $3\,n_0$, and the $\Xi^0$ hyperon at about $6\,n_0$. Hyperons $\Sigma^+$ and $\Sigma^0$ emerge only after $10\,n_0$. Due to the wide availability of redistribution channels in a wide range of densities, the adiabatic index $\Gamma_{\mathrm{EQ}}$ is significantly lower than $\Gamma_{\mathrm{FR}}$ from about $3\,n_0$ to beyond $10\,n_0$. As usual, $\Gamma_{\mathrm{FR}}$ shows a smoother and more gradual decline.

The behavior of $\omega_{\mathrm{EQ}}$ is very similar to that shown in Figure~\ref{Fig:multiple-gamma-frec}(e). However, the behavior of $\omega_{\mathrm{FR}}$ changes significantly compared to the cases without hyperons. 
The most notable change is that $\omega_{\mathrm{FR}}$ remains positive well beyond the maximum mass configuration because $\Gamma_{\mathrm{FR}}$ is significantly higher than $\Gamma_{\mathrm{EQ}}$ in a density range extending far beyond the central density of the maximum mass object ($\sim 6\,n_0$).
As a result, the extended branch in the frozen composition scenario is significantly longer than in cases without hyperons, as indicated by the distance between the circle and the cross on the curve $M$ versus $n_{b_c}$ in Figure~\ref{Fig:multiple-gamma-frec}(g). While the last stable configuration for equilibrium oscillations occurs at $n_{b_c} \sim 6\,n_0$, in the frozen composition scenario, the last dynamically stable configuration is reached at $n_{b_c} \sim 9\,n_0$.

%--------------------------------
\subsection{Model $N\!H\!\Delta2$}
%--------------------------------

In Figure~\ref{Fig:multiple-gamma-frec}(d), we present the results for the $N\!H\!\Delta$ case with $x_{\sigma \Delta}=1.25$. Due to the higher value of $x_{\sigma \Delta}$ compared to Figure~\ref{Fig:multiple-gamma-frec}(c), all four types of resonance $\Delta$ are now present. The $\Delta$ appear in the same order and within the same density range ($\sim 1.5-3.5\,n_0$) as in the $N\!\Delta2$ model shown in Figure~\ref{Fig:multiple-gamma-frec}(b), which also used $x_{\sigma \Delta}=1.25$.

Compared to the $N\!H\!\Delta1$ model in Figure~\ref{Fig:multiple-gamma-frec}(c), the emergence of the four resonances $\Delta$ causes the onset of the $\Lambda^0$ and $\Xi^-$ hyperons to shift to significantly higher densities.
This pattern of particle abundances makes the behavior of $\Gamma_{\mathrm{EQ}}$ very similar to that of the $N\!\Delta2$ model in Figure~\ref{Fig:multiple-gamma-frec}(b) up to $\sim 3.5\,n_0$. However, beyond $\sim 3.5\,n_0$ in Figure~\ref{Fig:multiple-gamma-frec}(d), the particles $\Lambda^0$, $\Delta^{++}$, $\Xi^-$, and $\Xi^0$ appear sequentially, allowing for further chemical redistribution as $n_b$ varies. This results in a substantial reduction in $\Gamma_{\mathrm{EQ}}$ compared to the $N\!\Delta2$ model shown in Figure~\ref{Fig:multiple-gamma-frec}(b).

At low central densities, it can be seen  from Figure~\ref{Fig:multiple-gamma-frec}(h) that $\omega_{\mathrm{EQ}}$ and $\omega_{\mathrm{FR}}$ are nearly identical, similar to what we observed in previous models. However, much like in the $N\!\Delta2$ case, $\omega_{\mathrm{EQ}}$ shows a noticeable drop around $n_{b_c} \sim 2\,n_0$, which corresponds to the density range where the sequential appearance of $\Delta$ resonances causes a significant softening of the EoS. In contrast, $\omega_{\mathrm{FR}}$ remains relatively stable and higher than $\omega_{\mathrm{EQ}}$ due to the stiffer behavior of matter under the assumption of frozen composition. As the central density approaches $n_{b_c} \sim 7 \,n_0$, $\omega_{\mathrm{EQ}}$ tends to zero, indicating the onset of dynamic instability at the maximum mass model. On the other hand, $\omega_{\mathrm{FR}}$ remains positive until much higher central densities, resulting in an extended branch of dynamically stable configurations when the composition is frozen. The curve $\omega_{\mathrm{FR}}$ ultimately reaches zero near $n_{b_c} \sim 9\,n_0$, as shown by the green cross on the curve $M$ versus $n_{b_c}$.

%-------------------------------------------------------------
\section{Summary and Discussion} 
%-------------------------------------------------------------
\label{sec:conclu}

This work examined the stability of NSs, with a particular focus on the possibility that their oscillations may occur out of chemical equilibrium. We considered three different baryonic compositions for the EoS: (i) nucleonic matter composed solely of protons and neutrons ($N$); (ii) nucleonic matter including $\Delta$ resonances ($N\Delta$); and (iii) the full baryonic octet supplemented by $\Delta$ resonances ($N\!H\!\Delta$). The leptonic sector always consists of electrons and muons. For scenarios involving $\Delta$ resonances, we investigated two different values of the coupling constant $x_{\sigma\Delta}$, namely $x_{\sigma\Delta}=0.9$ and $x_{\sigma\Delta}=1.25$.
Using these various compositions, we solved the TOV equations to construct stellar configurations in hydrostatic equilibrium. We then evaluated their stability by determining the fundamental mode frequency of small, adiabatic, radial oscillations. 
The frequency calculation is performed under two limiting regimes, depending on the relative values of the nuclear timescale $\tau_\mathrm{nuc}$ and the pulsation period $\mathcal{T}$ of the NS: one in which the system maintains continuous chemical equilibrium ($\tau_\mathrm{nuc} \ll \mathcal{T}$) and another in which the chemical composition remains frozen throughout the oscillation process ($\tau_\mathrm{nuc} \gg \mathcal{T}$).
As demonstrated in the appendix, the set of pulsation equations is formally identical in both cases, the only difference being the choice of the adiabatic index - $\Gamma_\mathrm{EQ}$ or $\Gamma_\mathrm{FR}$ - appropriate to each physical scenario.

The analysis of the adiabatic indices reveals a clear trend. When stellar matter remains in chemical equilibrium during perturbations, significant particle redistributions can occur within each fluid element, particularly in density regimes that facilitate the emergence of new particle species. These redistributions substantially lower the effective adiabatic index under equilibrium conditions, making the star’s response to perturbations softer. In contrast, if the stellar composition remains fixed during the oscillation process, the inability of particles to readjust their abundances with changing density leads to a stiffer response characterized by higher adiabatic indices. 

The distinction between equilibrium and frozen scenarios tends to vanish once small density changes no longer trigger substantial compositional rearrangements, which typically occur at sufficiently high densities. However, in models with more complex compositions—especially those that include multiple hyperons and resonances—large regions of the star’s interior allow the emergence of new particle species, even at high densities. As a result, for these more complex EoSs, the difference between $\Gamma_\mathrm{EQ}$ and $\Gamma_\mathrm{FR}$ becomes even more pronounced.
Notice that the discrepancy between the two adiabatic indices is strongly influenced by the value of $x_{\sigma \Delta}$, as this parameter determines the thresholds at which $\Delta$ resonances emerge.

In all the models considered, the lower value of $\Gamma_\mathrm{EQ}$ leads to lower oscillation frequencies, while the frozen composition scenario, with its higher $\Gamma_\mathrm{FR}$, yields stronger restoring forces and thus higher frequencies. The consequence of this distinction is evident in the location of the dynamic instability threshold. As expected, when perturbed matter is able to attain fast chemical equilibrium, the maximum-mass NS is the first configuration in the sequence to become unstable.
In contrast, if the composition remains fixed during pulsations, the onset of instability (i.e., the configuration with $\omega_0^2 = 0$) occurs beyond the maximum mass point. This shift creates an extended stable branch despite $\partial M / \partial n_{b_c}$ being negative. To the best of our knowledge, this is the first time that the existence of a sufficiently long branch of purely hadronic stable configurations beyond the maximum mass has been reported in the literature. This physical scenario is comparable to that of hybrid stars with a slow conversion regime between hadronic and quark matter at the discontinuity that separates both phases, which was first discussed in \cite*{Pereira:2017rmp} and implemented in several models by \citep{mariani:2019mhs, malfatti:2020dba, tonetto:2020dgm, rodriguez:2021hsw, curin:2021hsw, Goncalves:2022ios, Mariani:2022omh, Ranea:2022bou, lugones:2023ama, Ranea:2023auq, Ranea:2023cmr, Rau:2023tfo, Rau:2023neo, Rather:2024roo,Jimenez:2024htq}.

For EoSs with simpler compositions, the difference between the equilibrium and frozen scenarios is more modest. As the star’s central density increases, both adiabatic indices tend to converge, causing the equilibrium and frozen oscillation frequencies to approach one another. Under these conditions, the extended stable branch extends only slightly beyond the equilibrium maximum mass. 
That is, while the fundamental mode frequency is sensitive to the adiabatic index, the terminal configuration remains close to the mass peak because the differences between $\Gamma_{\rm{FR}}$ and $\Gamma_{\rm{EQ}}$ are not significant in the density range that dominates in the cores of NSs with $M \sim M_\mathrm{max}$.

In contrast, for EoSs with more complex compositions featuring multiple hyperons and/or resonances, the disparity between the two indices becomes more pronounced over a wide range of densities. 
A substantial increase in $\Gamma_{\rm{FR}}$ relative to $\Gamma_{\rm{EQ}}$ within the density regime typical of NS cores with $M \sim M_\mathrm{max}$ significantly elevates the frozen oscillation frequencies above their equilibrium counterparts for stars with masses near the maximum mass.  As a result, the stable branch under frozen conditions can extend far beyond the $M_\mathrm{max}$ configuration identified in the equilibrium scenario. 
The preceding discussion suggests that the compositional diversity of the inner cores of NSs plays a critical role in the existence of extended branches.

When a more realistic scenario is considered in which various nuclear reactions have different timescales, the situation becomes more nuanced. If certain reactions occur on timescales shorter than the oscillation period $\mathcal{T}$, their associated particle species can readily reach chemical equilibrium during the star’s pulsations, behaving as if $\tau_\mathrm{nuc} \ll \mathcal{T}$. On the other hand, for reactions with much longer timescales ($\tau_\mathrm{nuc} \gg \mathcal{T}$), the relevant particle abundances will remain essentially frozen over an oscillation period. In this intermediate regime, the effective adiabatic index $\Gamma$ would lie somewhere between the extremes represented by $\Gamma_\mathrm{EQ}$ and $\Gamma_\mathrm{FR}$.

Of particular importance are the reaction timescales of the particle species that emerge in the stellar core. These species, such as hyperons or $\Delta$ resonances, appear at higher densities and therefore have the potential to significantly influence the extended branch of the star. If the particles that trigger substantial compositional rearrangements possess relatively fast reaction channels, their ability to reach equilibrium states during oscillations could limit the extension of the stable branch. In contrast, if these key particles have predominantly slow reactions, their composition remains effectively frozen on the timescale of the star’s fundamental mode, leading to a scenario closer to the $\Gamma_\mathrm{FR}$ limit and allowing the extended branch to grow longer.

In essence, the actual length of the extended branch will depend critically on the interplay between the characteristic timescales of the reactions producing new particle species and the oscillation period of the star. By placing realistic constraints on these nuclear reaction timescales, one can gain a more accurate understanding of the stability landscape of neutron stars and the true extent to which their stable configurations may be prolonged beyond the conventional maximum mass threshold.

While our analysis points to the potential for extended stable branches beyond the conventional maximum mass configuration, it is important to acknowledge that these solutions may, in fact, be metastable rather than truly stable. A more comprehensive hydrodynamic treatment is needed to fully assess their resilience; however, this is beyond the scope of this work. Under such scrutiny, even relatively small yet sufficiently disruptive perturbations might trigger gravitational collapse, revealing whether these stars reside in a firmly established equilibrium state or merely hover on the brink of instability. By evaluating the energy barriers that must be overcome to destabilize these configurations, it could be possible to better determine the nature of the extended branch and refine the understanding of the conditions under which NSs can resist collapse.

\begin{acknowledgments}
MOC-P is a doctoral fellow at CONICET (Argentina). GL acknowledges the partial financial support from CNPq-Brazil (grant 316844/2021-7) and FAPESP-Brazil (grant 2022/02341-9). MOC-P, IFR-S, and MGO acknowledge UNLP and CONICET (Argentina) for financial support under grants G187 and PIP 0169.  IFR-S is also partially supported by PIBAA 0724 from CONICET. The contribution to this work of MGO and IFR-S was partially supported by the National Science Foundation (USA) under Grant PHY-2012152. 
\end{acknowledgments}

\appendix

%-------------------------------------------------------------
\section{Equations for radial adiabatic stellar pulsations out of chemical equilibrium}
\label{sec:appendix}
%-------------------------------------------------------------

\citet{Chandrasekhar:1964zz} derivation of relativistic stellar oscillation equations is based on the assumption that perturbations occur while maintaining chemical equilibrium (catalyzed matter). Subsequent investigations into pulsations out of chemical equilibrium \citep{Meltzer1966, Chanmugam1977, Gourgoulhon1995, Gondek:1997rpa} adopted these same equations, but simply substituted the equilibrium adiabatic index $\Gamma_\mathrm{EQ}$ with the non-equilibrium index $\Gamma_1$, making no other adjustments. \citet{Rau:2023neo} recently attempted to re-derive the pulsation equations in a similar setting but unfortunately introduced certain incorrect terms. Therefore, in this appendix, we re-derive the pulsation equations for the general case where chemical equilibrium is not instantaneously achieved during adiabatic perturbations.

%-------------------------------------------------------------------------------------
\subsection{Generalization of Equation (53) from Chandrasekhar (1964)}
%-------------------------------------------------------------------------------------

We consider a cold EoS of the form $P = P(\epsilon, Y_i)$, where $\epsilon$ denotes the energy density, and $Y_i$ represents the particle abundances. The Eulerian ($\delta$) and Lagrangian ($\Delta$) variations of the pressure are given by:
\begin{equation}
\delta P = \frac{\partial P}{\partial \epsilon} \delta \epsilon + \frac{\partial P}{\partial Y_i} \delta Y_i, \qquad    \Delta P = \frac{\partial P}{\partial \epsilon} \Delta \epsilon + \frac{\partial P}{\partial Y_i} \Delta Y_i,
\label{eq:A1}
\end{equation}
where repeated indices imply summation.  When chemical equilibrium is established on a timescale much shorter than the oscillation period, the Eulerian perturbation in composition, $\delta Y_i$, becomes negligible. In contrast, if chemical equilibrium is achieved over a timescale significantly longer than the oscillation period, fluid elements oscillate with a fixed composition, resulting in the vanishing of the Lagrangian composition change, $\Delta Y_i$.

Replacing Eq.~(39) of \citet{Chandrasekhar:1964zz}
\begin{equation}
\delta \epsilon = -\xi \frac{d\epsilon_0}{dr} - (\epsilon_0 + P_0)\frac{e^{\nu_0/2}}{r^2} \frac{\partial}{\partial r}\left(r^2 e^{-\nu_0/2}\xi\right),
\end{equation}
into Eq.~\eqref{eq:A1}, we obtain:
\begin{equation}
\delta P = -\frac{\Gamma_1 P_0}{\epsilon_0 + P_0} \xi \frac{d\epsilon_0}{dr} - \Gamma_1 P_0 \frac{e^{\nu_0/2}}{r^2} \frac{\partial}{\partial r}\left( r^2 e^{-\nu_0/2} \xi \right) + P_0 \beta_{Y_i} \delta Y_i.
\label{eq:A3}
\end{equation}
Here, $P_0$ and $\epsilon_0$ denote the unperturbed equilibrium quantities, and we define:
\begin{equation}
\beta_{Y_i} \equiv \left. \frac{\partial \ln P}{\partial Y_i} \right|_{\epsilon, Y_j \neq Y_i}.
\end{equation}

To eliminate $d\epsilon_0/dr$ from Eq.~\eqref{eq:A3}, we start with the relation $dP = (\partial P / \partial \epsilon) d\epsilon + (\partial P/\partial Y_i)  dY_i$ and use the fact that since $\xi$ is small and any thermodynamic quantity $X$ can be expressed as $X(r) = X_0(r) + \delta X$, the following approximation holds at first order: $\xi {d X_0}/{dr} \approx \xi {d X}/{dr}$.
Therefore, we find:
\begin{equation}
-\frac{P_0\Gamma_1}{\epsilon_0+P_0} \xi \frac{d\epsilon_0}{dr} = -\xi \frac{dP_0}{dr} + P_0\beta_{Y_i}\xi \frac{dY_{i0}}{dr},
\end{equation}
where $Y_{i0}$ represents the unperturbed configuration. When substituted into Eq.~\eqref{eq:A3}, this yields:
\begin{equation}
\delta P = -\xi \frac{dP_0}{dr} + P_0 \beta_{Y_i}\xi \frac{dY_{i0}}{dr} - \Gamma_1 P_0 \frac{e^{\nu_0/2}}{r^2}\frac{\partial}{\partial r}\left(r^2 e^{-\nu_0/2}\xi\right) + P_0 \beta_{Y_i} \delta Y_i.
\end{equation}

Using the relationship between the Eulerian and Lagrangian variations of the chemical composition, \mbox{$\Delta Y_i = \delta Y_i + \xi {dY_{i0}}/{dr}$}, we obtain the generalized form of Chandrasekhar’s Eq.~(53):
\begin{equation}
\Delta P = \delta P +  \xi \frac{dP_0}{dr}  = - \Gamma_1 P_0 \frac{e^{\nu_0/2}}{r^2}\frac{\partial}{\partial r}\left(r^2 e^{-\nu_0/2}\xi\right) + P_0 \beta_{Y_i}\Delta Y_i.
\label{eq:A7}
\end{equation}
In this equation, in addition to replacing $\Gamma_\mathrm{EQ}$ with $\Gamma_1$, extra composition-dependent terms appear compared to the original equation. If the composition is frozen, the last term vanishes, and the equation reduces to Chandrasekhar’s Eq.~(53), with $\Gamma_1$ substituting $\Gamma_\mathrm{EQ}$.

%-------------------------------------------------------------------------------------
\subsection{Generalization of Equation (55) from Chandrasekhar (1964)}
%-------------------------------------------------------------------------------------

Following Section VI of \citet{Chandrasekhar:1964zz}, we substitute $\delta \epsilon$ from his Eq.~(37) and our generalized expression for $\delta P$ into his Eq.~(43). This procedure leads to a generalized form of his Eq.~(55):
\begin{equation}
\begin{aligned}
\omega^2 e^{\lambda_0-\nu_0}(P_0+\epsilon_0)\xi =
-\frac{d}{dr}\left(\xi\frac{dP_0}{dr}\right)
-\left(\tfrac{1}{2}\frac{d\lambda_0}{dr}+\frac{d\nu_0}{dr}\right)\xi\frac{dP_0}{dr}
-\tfrac{1}{2}(P_0+\epsilon_0)\left(\frac{d\nu_0}{dr}+\frac{1}{r}\right)\left(\frac{d\lambda_0}{dr}+\frac{d\nu_0}{dr}\right)\xi \qquad \\
-\tfrac{1}{2}\frac{d\nu_0}{dr}\left\{\frac{\partial}{\partial r}[(P_0+\epsilon_0)\xi]+\frac{2}{r}(P_0+\epsilon_0)\xi\right\}
- e^{-(\lambda_0/2+\nu_0)} \frac{d}{dr}\left[e^{\lambda_0/2+\nu_0}\left(\frac{\Gamma_1 P_0 e^{\nu_0/2} }{r^2}\frac{d}{dr}(r^2 e^{-\nu_0/2}\xi)  - P_0\beta_{Y_i}\Delta Y_i\right)\right].
\end{aligned}
\end{equation}
Following a similar derivation process as in Section VI of Chandrasekhar (1964), one also obtains the generalized form of his Eq.~(59), known as the ``pulsation equation'':
\begin{equation}
\begin{aligned}
\omega^2 e^{\lambda_0-\nu_0}(P_0+\epsilon_0)\xi &= \frac{4}{r}\frac{dP_0}{dr}\xi + 8\pi e^{\lambda_0} P_0 (P_0+\epsilon_0)\xi - \frac{1}{P_0+\epsilon_0}\left(\frac{dP_0}{dr}\right)^2\xi \\
&\quad + e^{-(\lambda_0/2+\nu_0)}\frac{d}{dr}\left[e^{\lambda_0/2+\nu_0}\left(-\frac{\Gamma_1 P_0}{r^2}e^{\nu_0/2}\frac{d}{dr}(r^2 e^{-\nu_0/2}\xi) + P_0 \beta_{Y_i}\Delta Y_i\right)\right].
\end{aligned}
\label{eq:A9}
\end{equation}
The key modifications relative to Chandrasekhar’s original equation are the inclusion of a composition-dependent term and the use of $\Gamma_1$ instead of the original adiabatic index $\Gamma_\mathrm{EQ}$.

%-------------------------------------------------------------------------------------
\subsection{Pulsation equations in the form used in this work}
%-------------------------------------------------------------------------------------

In Chandrasekhar’s original formulation, the variable $\xi = \Delta r$ denotes the radial displacement. Here, we redefine the variable so that it represents the relative radial displacement:
\begin{equation}
\xi = \frac{\Delta r}{r}.
\end{equation}
The second-order pulsation equation can be transformed into a system of two first-order equations in terms of the variables $\Delta P$ and the newly defined $\xi$. This approach, proposed by \citet{Gondek:1997rpa}, is particularly convenient for numerical implementations. 
For simplicity, we will omit the subscript $0$ from the unperturbed quantities.

The first pulsation equation is obtained by expanding the radial derivative in Eq.~\eqref{eq:A7} and rearranging the terms:
\begin{equation}
\frac{d \xi}{d r} = -\frac{1}{r}\left(3 \xi - \frac{\Delta P}{P \Gamma_1}\right) - \frac{dP}{dr} \frac{\xi}{(\epsilon + P)} + \frac{\beta_{Y_i} \Delta Y_i}{r \Gamma_1}.
\end{equation}
Compared to the equation in \citet{Gondek:1997rpa} for catalyzed matter, this expression introduces an additional composition-dependent term and substitutes $\Gamma_\mathrm{EQ}$ with $\Gamma_1$. Since this work focuses on oscillations with frozen composition, the last term of the above equation is neglected in Eq. \eqref{Eq:ecuacionparaXI}.

The second pulsation equation is derived by recognizing that the right-hand side of Eq.~\eqref{eq:A7} is present in the final term of Eq.~\eqref{eq:A9}. By substituting this expression with $\Delta P$ and expanding the derivative, one obtains:
\begin{equation}
\begin{aligned}
\omega^2 e^{\lambda-\nu}(P+\epsilon) r \xi &= 4 \xi \frac{d P}{d r}  + 8\pi e^{\lambda}P(P+\epsilon) r \xi - \frac{1}{(P+\epsilon)}\left(\frac{d P}{d r}\right)^2 r \xi + \frac{d}{d r}(\Delta P) + \Delta P \left(\tfrac{1}{2}\frac{d \lambda}{d r}+\frac{d \nu}{d r}\right).
\end{aligned}
\end{equation}
Using Eq.~(22) of \citet{Chandrasekhar:1964zz}, we can replace the factor ($\tfrac{1}{2}d\lambda/dr+ d\nu/dr$) in the previous equation, which results in:
\begin{equation}
\begin{aligned}
\frac{d}{d r}(\Delta P) &= \xi  \left[\omega^2 e^{\lambda-\nu}(P+\epsilon)r - 4 \frac{d P}{d r}\right] + \xi  \left[\frac{r}{(P+\epsilon)}\left(\frac{d P}{d r}\right)^2 - 8\pi e^{\lambda}(P+\epsilon)P r\right] \\
&\quad + \Delta P \left(\frac{1}{(P+\epsilon)} \frac{d P}{dr} - 4\pi(P+\epsilon) r e^{\lambda}\right).
\end{aligned}
\label{eq:A13}
\end{equation}
This equation matches the one derived by \citet{Gondek:1997rpa}, meaning that it retains exactly the same form for both catalyzed and non-catalyzed matter.

\end{document}